\documentclass[a4paper,fleqn,usenatbib,useAMS]{mnras}

\usepackage{newtxtext,newtxmath}
\usepackage{natbib}
\usepackage[T1]{fontenc}

\DeclareRobustCommand{\VAN}[3]{#2}
\let\VANthebibliography\thebibliography
\def\thebibliography{\DeclareRobustCommand{\VAN}[3]{##3}\VANthebibliography}

\usepackage{graphicx}	
\usepackage{amsmath}	
\usepackage{subfigure}

\usepackage[dvipsnames]{xcolor}
\definecolor{blazeorange}{rgb}{1.0, 0.4, 0.0}
\definecolor{seagreen}{rgb}{0.18, 0.55, 0.34}
\definecolor{rufous}{rgb}{0.66, 0.11, 0.03}
\definecolor{royalfuchsia}{rgb}{0.79, 0.17, 0.57}
\definecolor{scarlet}{rgb}{1.0, 0.13, 0.0}
\definecolor{royalpurple}{rgb}{0.47, 0.32, 0.66}
\usepackage{ulem} 

\title[Optical Counterparts of Two Candidate ULXs]{Optical Counterparts of Two Candidate Ultraluminous X-ray Sources in NGC\,4536}

\author[H. Avdan et al.]{
H. Avdan,$^{1}$\thanks{E-mail: avdan.hsn@gmail.com}
E. Sonbas,$^{1,2}$
K. S. Dhuga,$^{2}$
A. Vinokurov,$^{3}$
E. G\"o\u{g}\"u\c{s},$^{4}$
S. Avdan,$^{1}$
\newauthor Y. N. Solovyeva,$^{3}$ 
A. E. Kostenkov,$^{3}$
E. S. Shablovinskaya,$^{3}$
and D. Goktas$^{1}$ \\
\\
$^{1}$Adiyaman University, Department of Physics, 02040 Adiyaman, Turkey\\
$^{2}$Department of Physics, The George Washington University, Washington, DC 20052, USA\\
$^{3}$Special Astrophysical Observatory of the Russian AS, Nizhnij Arkhyz, Russia\\
$^{4}$ Faculty of Engineering and Natural Sciences, Sabanc\i~University, Orhanl\i~- Tuzla, Istanbul 34956, Turkey
}

\date{Accepted XXX. Received YYY; in original form ZZZ}

\pubyear{2015}

\begin{document}
\label{firstpage}
\pagerange{\pageref{firstpage}--\pageref{lastpage}}
\maketitle

\begin{abstract}
Archival {\it XMM-Newton}, {\it Chandra} and {\it Hubble Space Telescope (HST)} data have been used to study the X-ray and optical properties of two candidate ultraluminous X-ray sources (ULXs) in NGC\,4536. In order to search for potential optical counterparts, relative astrometry between {\it Chandra} and {\it HST} was improved, and as a result, optical counterparts were detected for both X-ray sources. To complement our findings (based on the archival data), ground-based optical spectra of the counterparts were obtained with the 6m BTA Telescope located at the Special Astrophysical Observatory (SAO). The calculated redshift (z = 0.4391$\pm$0.0010) for one of the sources (X-3) indicates that the source is, in fact, a background active galactic nucleus (AGN). Two possible optical counterparts (s1 and s2) were found for X-2. Whether s1 is point-like or an extended source is unclear: If it is point-like and the emission is dominated by the donor its spectral type indicates O-B star. The second source (s2) is point-like and is consistent with the colors and absolute magnitudes of a red supergiant. 

\end{abstract}




\section{Introduction}

Ultraluminous X-ray sources (ULXs, with a threshold luminosity of $L_{\mathrm{X}} > 10^{39}$ erg~s$^{-1}$), are non-nuclear sources that have been found in diverse environments, including star-forming regions of spiral galaxies, elliptical galaxies, as well as dwarf galaxies \citep{Feng2011,Kaaret2017,Fabrika2021}. Assuming isotropic emission, some of these sources can reach a luminosity well in excess of $L_{\mathrm{X}} > 10^{40}$ erg~s$^{-1}$, surpassing the Eddington limit for a~10$M_{\sun}$ black hole (BH) \citep{Colbert1999,Farrell2009,Sutton2012}. Early studies hinted that the majority of ULXs as being binary systems powered by accretion onto a compact object where the compact object could be either an intermediate mass ($10^{2}-10^{4} M_{\sun}$) or a stellar mass BH ($\le 100 M_{\sun}$), accreting at sub-, near or above the Eddington rate \citep{Colbert1999,King2001,Begelman2002,Roberts2007,Tao2011,Fabrika2015,Kaaret2017}.  More recent studies, with detection of pulsations in a handful of sources \citep{Bachetti2014,Furst2016,Israel2017,Carpano2018,Sath2019,Rodri2021,Quintin2021}, strongly argue in favor of (at least) a fraction of these sources hosting a neutron star (NS) instead of a BH. With the current total of NS ULXs (with observed pulsations) standing at about a half-dozen, the question of the relative proportion of ULXs hosting a NS, as opposed to a stellar-mass BH, has become a highly debated topic \citep{Middleton2017,King2016}.\\

\noindent ULXs appear to exhibit different spectral and timing behavior in comparison with (ordinary) galactic BHs, featuring two-component spectra with soft excess and a turnover at energies near 5 keV. As well, in contrast to galactic BH binaries, ULXs tend to be more persistent rather than transient. Moreover, ULXs found in elliptical galaxies tend to have lower levels of variability than those found in star-forming galaxies, that can vary by an order of magnitude \citep{Feng2006,feng2009,Soria2012,Pintore2018,Atapin2020,Walton2021}. Indeed, \citet{Sutton2013} extracted fractional variability and constructed variability-hardness diagrams to distinguish three main states i.e., the broadened disk (BD: possibly dominated by a slim disk, overall luminosity $\sim 3 \times 10^{39}$ erg s$^{-1}$), hard ultraluminous (HUL; low-inclination i.e., more face-on; luminosity $L_{\mathrm{X}} > 3 \times 10^{39}$ erg s$^{-1}$), and the soft ultraluminous (SUL; high inclination i.e, more edge-on; luminosity $L_{\mathrm{X}} > 3 \times 10^{39}$ erg s$^{-1}$). Variabilities reaching $25-40 \%$ have been reported for the SUL and BD states. \citet{Kajava2009} have noted that the 'soft excess' reported for many ULXs around 0.2 keV does not follow the expected temperature profile ($L \propto T^{4}$) of a standard disk. These considerations raise very important questions regarding not only the configuration (and/or the contribution) of the accretion disk but the accretion process itself i.e, whether it is primarily disk-fed or driven via clumpy winds or possibly even a mixture of the two with the development of transient disk structures.\\

\noindent Optical observations of the environments of ULXs provide important information regarding the nature of these sources and also the evolution patterns of X-ray binaries \citep{Poutanen2013,Fabrika2021,Kaaret2017}. The optical emissions detected in ULXs could arise from an accretion disk, as a result of irradiation, and/or from the donor star \citep{Feng2011,Tao2011}. Determining the main source of the emissions would help in estimating the mass of the compact object and, therefore, provide a better understanding of the physical nature of ULXs. Many studies have shown that ULXs are associated with crowded star formation regions in their host galaxies \citep{Ranalli2003,Grimm2003,Kaaret2004,Swartz2009}. Therefore, to identify counterparts in such cases, the high spatial resolution and superior pointing capacity of Chandra coupled with the excellent resolution and imaging capacity of the {\it Hubble Space Telescope (HST)}, provide a powerful combination for precise astrometry \citep{Patruno2010,Grise2011,Poutanen2013,Avdan2016,Egorov2017,Urquhart2018,Avdan2019}. The number of ULXs defined by optical observations is still limited since they are quite faint in the optical band ($m_{V} > 21$ mag). With the aid of recent {\it HST} data the optical counterpart candidates for a number of ULXs have been studied in detail \citep{Tao2011, Soria2012, Gladstone2013,Avdan2016,Avdan2019,Akyuz2020} and O-B type (and less commonly cold) stars have been identified as unique optical counterparts for over a dozen ULXs  \citep{Avdan2019,Liu2004,Kuntz2005,Grise2012,Gladstone2013,Vinokurov2018}. Several studies, based on ground-based observations, have reported the optical spectra of a number of ULXs (or their environment) \citep{Motch2011, Avdan2016, Vinokurov2018, Fabrika2015}. These studies found that the observed spectra show similar structures in which they exhibit broad emission lines (e.g. He II $\lambda4686$) \citep{Vinokurov2018,Fabrika2021}. In some cases where background active galactic nuclei (AGNs) are misidentified and catalogued as ULXs, ground-based spectral observations are crucial in distinguishing between them \citep{Masetti2003,Arp2004,Clark2005,Dadina2013,Gutierrez2013,Lopez2020,Vinokurov2020}.\\

\noindent Our prime focus in this paper is the multiwavelength study of two ULX candidates (X-2 and X-3) in NGC\,4536, a late-type barred spiral galaxy at a distance of 15.8 Mpc in the Virgo constellation \citep{Mcalphine2011,Earnshaw2019,Hatt2018,deVaucouleurs1976}. \citet{Mcalphine2011} studied this galaxy with the aim of probing the nature of the nuclear and off-nuclear X-ray emission using {\it XMM-Newton}, supplemented with  optical data from {\it HST}, infrared imaging data from the {\it Spitzer} and {\it GALEX} (Galaxy Evolution Explorer). They detected statistically significant variability which provides evidence of an AGN. They also explored the possibility of X-ray emission due to existence of X-ray binaries (XRBs) and found that XRBs could be responsible for only a small fraction of the X-ray luminosity detected, and that the significant [Ne\,v] emission in the galaxy cannot be explained by the presence of XRBs alone. \citet{Singha2019} performed a spectral and temporal analysis of ULXs in NGC\,4536 and identified three X-ray point sources (X-ray counts $>$ 100 counts/sec) and concluded that these sources are ULXs with X-ray luminosity of $>$ $10^{39}$~erg\,s$^{-1}$. \citet{Earnshaw2019} catalogued three sources as ULXs in NGC\,4536 galaxy. In this study, we investigate two of them since the third one was not observed with {\it HST}.  Our goal is to search for and identify potential optical counterparts for these ULX candidates using the archival {\it Chandra} and {\it HST} observations. For completeness the X-ray spectra of the candidate ULXs were also obtained using {\it Chandra} and {\it XMM-Newton} archival data.\\

\noindent The paper is organized as follows: Section 2 reports the details of the observations, the data reduction, and methodology; Section 3 presents the results of the optical and X-ray spectral analysis. In section 4, we summarize our findings.

\section{Data Reduction and Methodology}
\subsection{Observations}
NGC\,4536 was observed with {\it XMM-Newton} and {\it Chandra} for $\sim$ 34ks (in 2008) and  $\sim$ 15ks (2017), respectively, and with {\it HST} in 2020. The details of the observations used in this work are summarized in Table~\ref{tab:obs_log}.\\

\begin{figure*}
\begin{center}
\includegraphics[angle=0,scale=0.268]{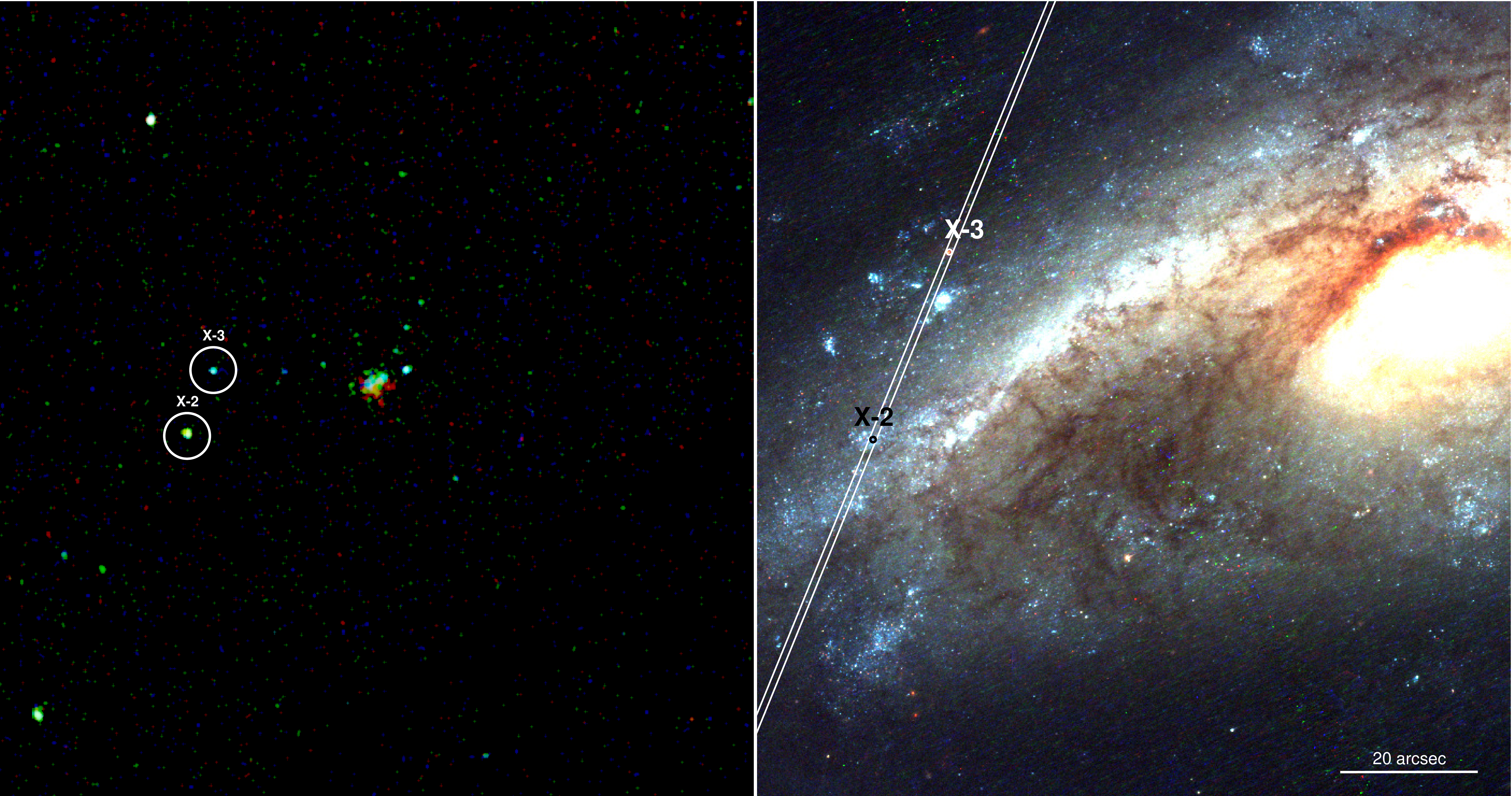}
\caption{ RGB X-ray {\it Chandra} (left) and {\it HST} (right) images of the NGC\,4536 galaxy. For the X-ray image, Red, Green and Blue represent 0.3–1 keV, 1–2.5 keV and 2.5–8 keV emissions, respectively and the images are smoothed with a 3$\arcsec$ Gaussian. For the {\it HST} RGB image, Red, Green and Blue represent F814W, F555W and F438W filters, respectively. North is up and east is to the left. The sources are indicated with black and white circles. The white parallel lines on the {\it HST} image depict the (1$\arcsec$ wide) slit position used during the BTA observation.}
\label{fig:ulxrgb}
\end{center}
\end{figure*}

\noindent The {\it XMM-Newton} data are used for spectral analysis because the sources do not have enough statistics in the {\it Chandra} observation. The archival {\it Chandra} observation is used to correct relative astrometry between {\it HST} and {\it Chandra} images in order to search for possible optical counterparts of the ULX candidates (see Figure~\ref{fig:ulxrgb} for the  location of the sources). The optical photometry is performed with the archival {\it HST}/WFC3/UVIS observations.\\

\noindent Both ULX candidates (X-2 and X-3) were also observed on 2021 May 13 using the 6-m Big Telescope Alt-azimuth (BTA) of the Special Astrophysical Observatory (SAO) with a SCORPIO-2 spectrograph \citep{Afanasiev2011} in the long slit mode. A total of 4 exposures of 1200 s were taken. The selected grism VPHG940@600 and slit width of 1\arcsec\ provided a spectral resolution of $\approx6$\,\AA\ in the range of $\approx3500 - 8500$\,\AA. The slit position (PA\,=\,157.5$^\circ$) during the observation is shown in the right panel of Figure~\ref{fig:ulxrgb}.

\begin{table}
	\centering
	\caption{{\it XMM-Newton}, {\it Chandra} and {\it HST} archival data used in this work.}
	\label{tab:obs_log}
	\begin{tabular}{lccr} 
		\hline
  Satellite   & ObsID (Filter)            & Date          & Exp. \\
                \hline

{\it XMM-Newton}      & 0551450301        & 2008 June 17  &   34 ks   \\

\hline
{\it Chandra}         & 19387             & 2017 July 01  &   15 ks   \\
\hline
{\it HST} (WFC3/UVIS) &  IDXR29060 (F275W) & 2020 March 22 & 2190 s \\
                &  IDXR29050 (F336W) & 2020 March 22 & 1100 s \\
                &  IDXR29040 (F438W) & 2020 March 22 & 1050 s \\
                &  IDXR29070 (F555W) & 2020 March 22 & 670 s  \\
                &  IDXR29030 (F814W) & 2020 March 22 & 803 s\\
                \hline

	\end{tabular}
\end{table}

\subsection{Astrometry and Optical Photometry}
\label{sec:phot}
\begin{figure}
\begin{center}
\includegraphics[angle=0,scale=0.5]{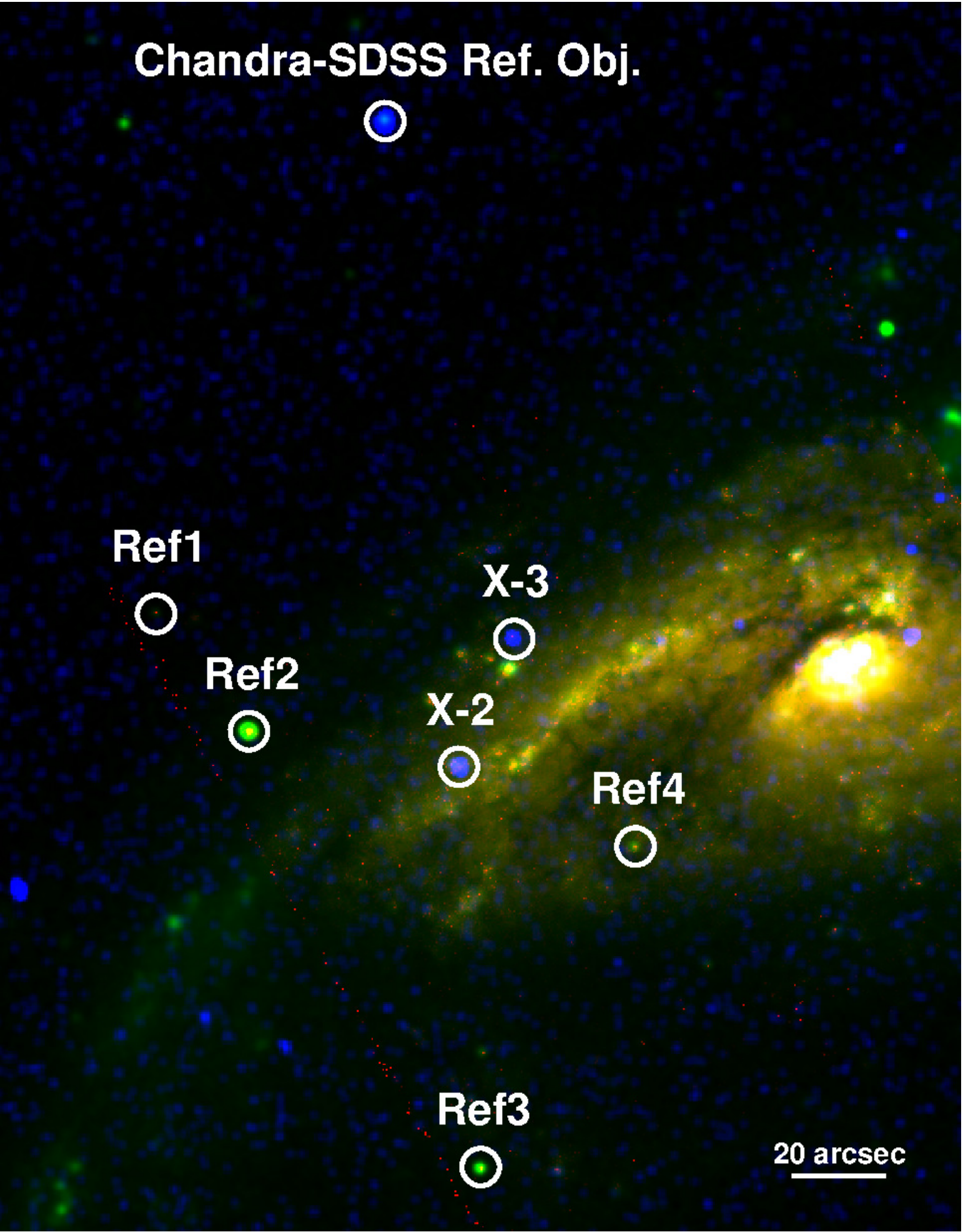}
\caption{Three-color composite image of NGC 4536 using the {\it HST}/WFC3 F555W (red), SDSS-g band (green), and {\it Chandra} (blue) images. The white circles represent the reference objects used in the astrometric corrections along with the positions of the sources X-2 and X-3. North is up and east is to the left.}
\label{fig:references}
\end{center}
\end{figure}

\begin{figure*}
\centering
\includegraphics[angle=0,scale=0.9]{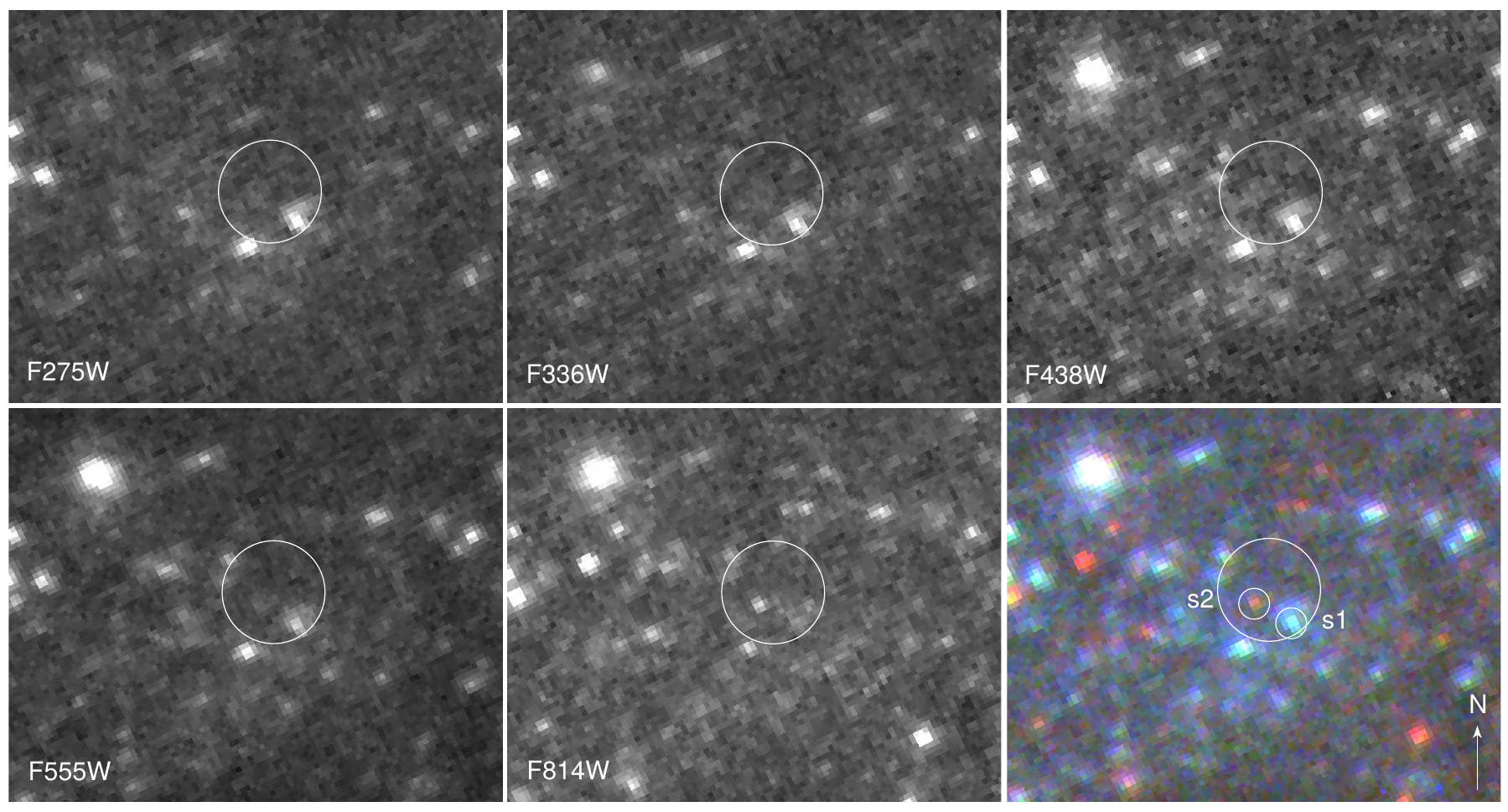}
    \caption{The archival {\it HST}/WFC3 images of the optical counterpart candidates for X-2. White circle has a radius of 0.4\arcsec and represents the uncertainty of the astrometric correction. The three-color image was obtained by taking red, green and blue images as F814W, F555W and F438W filters, respectively.}
    \label{fig:ulx2_hst}
\end{figure*}

\begin{figure*}
\includegraphics[angle=0,scale=0.9]{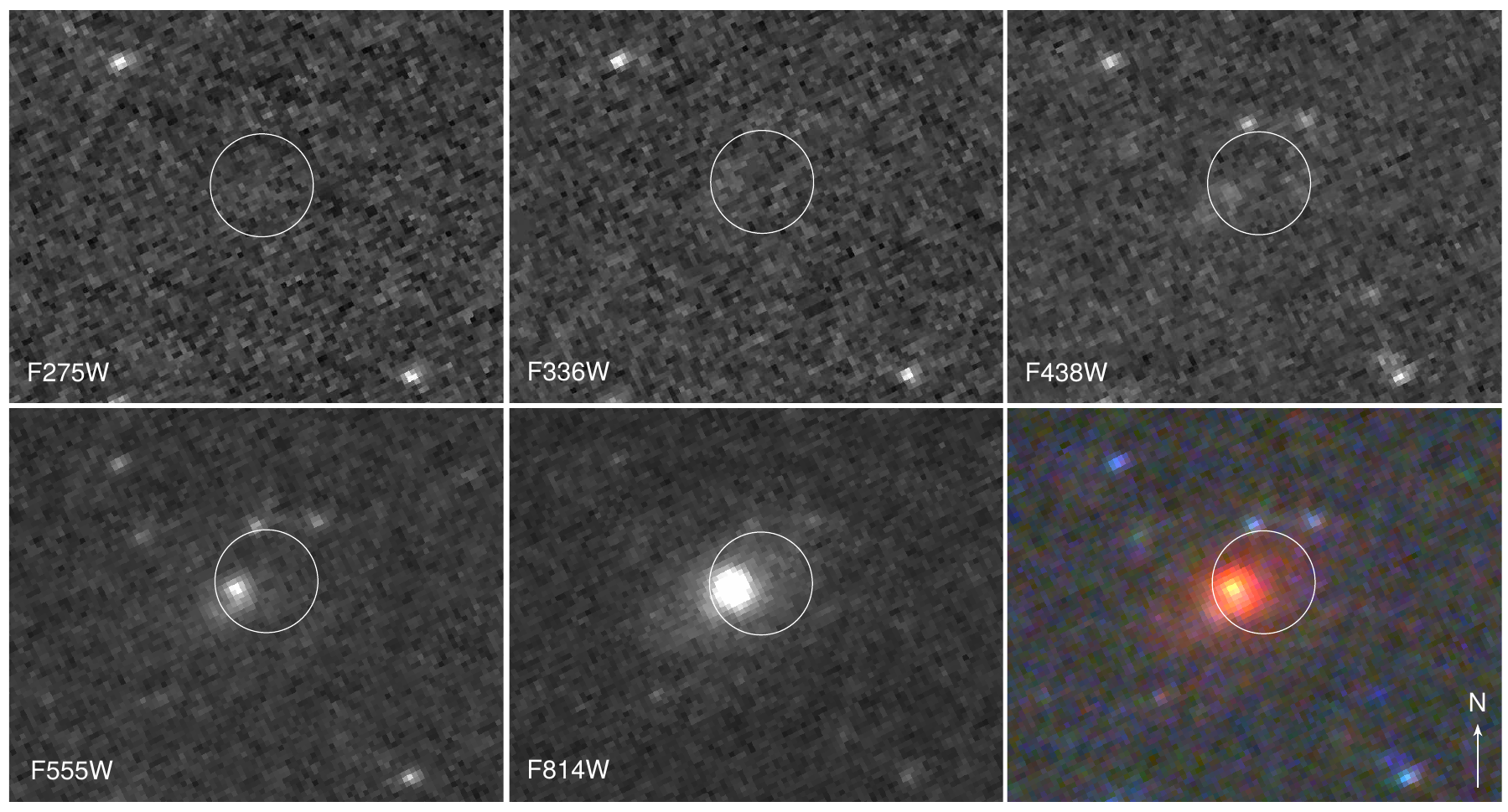}
    \caption{The archival {\it HST}/WFC3 images of the optical counterpart candidate for X-3. White circle has a radius of 0.4\arcsec and represents the uncertainty of the astrometric correction. The three-color image was obtained by taking red, green and blue images as F814W, F555W and F438W filters, respectively.}
    \label{fig:ulx3_hst}
\end{figure*}

A relative astrometric correction is required to find the accurate positions of the candidate sources in the {\it HST} images. Therefore, {\it Chandra} and {\it HST}/WFC3 images were used to apply astrometric corrections by using the methods described in (\citealt{Vinokurov2018} and references therein). As there were not a sufficient number of common sources between the {\it Chandra} and {\it HST} images, we tried to tie the images via an intermediate step. The SDSS g-band image of the galaxy was used as a reference image to determine the shift between the {\it Chandra} and {\it HST}/WFC3 F555W images. First the {\it Chandra} and {\it HST} images were aligned with the SDSS image individually. One bright point-like source was used to calculate the shift between the {\it Chandra} and the SDSS images and four bright sources were used for the {\it HST} and the SDSS images. The list of these reference objects, the original position of ULXs in {\it Chandra} images and the final corrected positions on the {\it HST}/WFC3 F555W image are given in Table~\ref{tab:ref_list}. The positions of the reference objects and ULXs (X-2 and X-3) on the three-color composite image are also shown in Figure~\ref{fig:references}. After the correction, the positions of X-2 and X-3 were derived with an uncertainty of 0\farcs4 as RA, Dec; $12^{h}$$34^{m}$$32^{s}.554$, $+$2$\degr$10\arcmin54\farcs84 and $12^{h}$$34^{m}$$31^{s}.803$ $+$2$\degr$11\arcmin22\farcs23 on the {\it HST}/WFC3 F555W image, respectively. The {\it HST}/WFC3 images of the sources are given in Figures~\ref{fig:ulx2_hst} and \ref{fig:ulx3_hst} respectively.\\

\begin{table}
	\centering
	\caption{The reference objects used to match {\it HST}$-$SDSS and {\it Chandra}$-$SDSS images, the original positions of the ULXs in {\it Chandra} image and the final corrected positions of the ULXs in {\it HST}/WFC3 F555W filter image. }
	\label{tab:ref_list}
	\begin{tabular}{lcccc} 
		\hline
Object & Image   & RA            & Dec          & Difference \\
        &         & (J2000)       & (J2000)      & ($\arcsec$) \\

\hline
\multicolumn{5}{c}{{\it HST}/WFC3 F555W $-$ SDSS-g band} \\
\hline

Ref1 &  {\it HST}          &  12 34 36.883                &   +02 11 27.359          &    0.07      \\
    &    SDSS             &  12 34 36.881                &   +02 11 27.428          &             \\

Ref2 &  {\it HST}          &  12 34 35.547                &   +02 11 02.005          &    0.16      \\
 &    SDSS             &  12 34 35.553                &   +02 11 02.137          &             \\
   
Ref3 &  {\it HST}          &  12 34 32.263                &   +02 09 29.169          &    0.17      \\
    &    SDSS             &  12 34 32.269                &   +02 09 29.311          &             \\

Ref4 &  {\it HST}          &  12 34 30.073                &   +02 10 37.472          &      0.21    \\
    &    SDSS             &  12 34 30.082                &   +02 10 37.635          &             \\

\hline
\multicolumn{5}{c}{{\it Chandra} $-$ SDSS-g band} \\
\hline

Ref  &  {\it Chandra}      &  12 34 33.632                & +02 13 11.676            &     0.49     \\
    &    SDSS             & 12 34 33.627                 & +02 13 12.162            &             \\

\hline
\multicolumn{5}{c}{ULXs} \\
\hline

X-2  & {\it Chandra}                         & 12 34 32.563              & +02 10 54.484    & 0.39                   \\
    & Corrected   & 12 34 32.554             & +02 10 54.844    &                   \\

X-3  & {\it Chandra}                         & 12 34 31.812              & +02 11 21.867    &     0.39              \\
 & Corrected   & 12 34 31.803              & +02 11 22.227   &                   \\

\hline

\end{tabular}
\end{table}


\noindent We have identified two relatively bright sources within the error box as possible optical counterparts of X-2 (marked as s1 and s2 in the color image of the Figure~\ref{fig:ulx2_hst}). An estimate of the size of the blue source s1, using a 2-D Gaussian function in F275W, F336W and F438W images, gives an average size (FWHMs) of $0.122\arcsec \times 0.096\arcsec$, which exceeds the size of the surrounding stars whose FWHM equals 0.082\arcsec with a standard deviation of less than 0.007\arcsec. However, the broadening may be due to a compact structure that is visible in the images to the north of s1 and is poorly spatially resolved from s1. So, we cannot make an unambiguous conclusion whether s1 is an extended source (for example, the core of a young star cluster) or a point-like source\footnote{Note that in the PSF photometric results, DOLPHOT gives a flag parameter as an indicator for the source type. The software returned a flag value of “1” for s1 indicating it to be a point-like source with a $\chi{^2}$ PSF fit of 1.28 and 1.01 in the filters F275W and F336W respectively.}. The other optical counterpart of X-2, i.e., s2, is a red point-like source clearly visible only in F555W and F814W images. Within the positional error box for X-3 there is a single extended (FWHM\,=\,$0.18\arcsec \times 0.16\arcsec$) optical source whose morphology and color indicates that it is a high redshift galaxy.\\

\noindent For s1 and s2 we performed point spread function (PSF) and aperture photometry and compared the results. PSF photometry was carried out for the {\it HST}/WFC3 FLT images using {\scshape dolphot} software version 2.0 \citep{Dolphin2000,Dolphin2016}. The WFC3/F555W drizzled image (DRZ) was chosen as the positional reference image for source detection. An aperture photometry was done for the WFC3 images in DRC format using {\scshape apphot} package in {\scshape iraf}. We chose a small aperture radius of 2 pixels ($\approx 0.08\arcsec$) because of the crowded field; aperture corrections were calculated using magnitudes of $3-9$ (depending on the filter) isolated bright stars measured in 2 and 10 pixels apertures. WFC3/UVIS filters' zero points for the 10 pixels aperture were taken from the uvis-photometric-calibration webpage\footnote{https://www.stsci.edu/hst/instrumentation/wfc3/data-analysis/photometric-calibration/uvis-photometric-calibration/previous-uvis-photometric-calibrations}. The background level was estimated in the annulus around the sources with various inner and outer radii from a few pixels to $r_{in} = 20$ and $r_{out} = 40$ pixels, for the aperture stars only large annulus was used. For the s1 source, aperture photometry always gave measurements up to 0.5\,mag brighter than those obtained by PSF photometry. The smallest difference (0.1–0.2 mag) between the results was obtained when the background was subtracted using a very small annulus with $r_{in} = 3$ and $r_{out} = 4$ pixels. This indicates a rather strong influence of a compact stellar environment of the source s1, whose contribution is optimally taken into account only when performing PSF photometry. The magnitudes of s2 remain constant within the errors and are in good agreement with the PSF measurements, regardless of the size of the annulus chosen to estimate the background level. These results make PSF measurements preferable for both for s1 and s2. Unlike the optical sources within the X-2 error box, there is little doubt that the unique optical counterpart candidate in the X-3 error box is point-like. Therefore, only aperture photometry was carried out for this source. The source magnitudes were measured in a large aperture with a radius of 10 pixels, the background was estimated in an annular aperture with $r_{in} = 20$ and $r_{out} = 40$ pixels. The reddening corrected instrumental VEGA magnitudes of the counterpart candidates with statistical $1\sigma$ errors are given in Table~\ref{tab:opt-photometry}. The correction was performed in the PySynphot package with adopted values E(B-V)=0.13 for X-2 (see section \ref{sec:spec} for details) and E(B-V)=0.017 (Galactic value, \citep{Schlafly2011}) for X-3.

\begin{table}
\centering
\caption{The reddening corrected instrumental VEGA magnitudes of the optical counterpart candidates. Adopted extinction values are A$_V = 0.41$\,mag for X-2 and  A$_V = 0.05$\,mag for X-3.}
\label{tab:opt-photometry}
\begin{tabular}{lccc}
\hline\hline
Filter & \multicolumn{2}{c}{X-2} & X-3 \\
\cline{2-3}
       &     s1     &      s2      &       \\
\hline
F275W  & $22.16\pm0.06$ & ---- & $24.4\pm0.3$ \\
F336W  & $22.94\pm0.07$ & ---- & $23.77\pm0.18$ \\
F438W  & $24.50\pm0.07$ & ---- & $24.55\pm0.10$ \\
F555W  & $24.77\pm0.08$ & $26.25\pm0.26$ & $23.023\pm0.023$ \\
F814W  & $25.51\pm0.23$ & $24.37\pm0.09$ & $20.491\pm0.008$ \\
\hline
\end{tabular}
\end{table}

\subsection{Optical Spectroscopy}
\label{sec:spec}
Spectroscopy was carried out on 2021 May 13 under relatively bad weather conditions, haze (clouds) appeared periodically, seeing was around 2\farcs2. Data processing was performed with the context {\scshape long} of {\scshape midas} using the standard algorithm, including bias subtraction, flat field correction, cosmic ray rejection, wavelength and flux calibration. Only the extraction of spectra was carried out without using the {\scshape midas} procedures, instead we used specialized software: {\scshape spextra} \citep{Sarkisyan2017}.\\

\begin{figure*}
\begin{center}
\includegraphics[angle=0,scale=0.65]{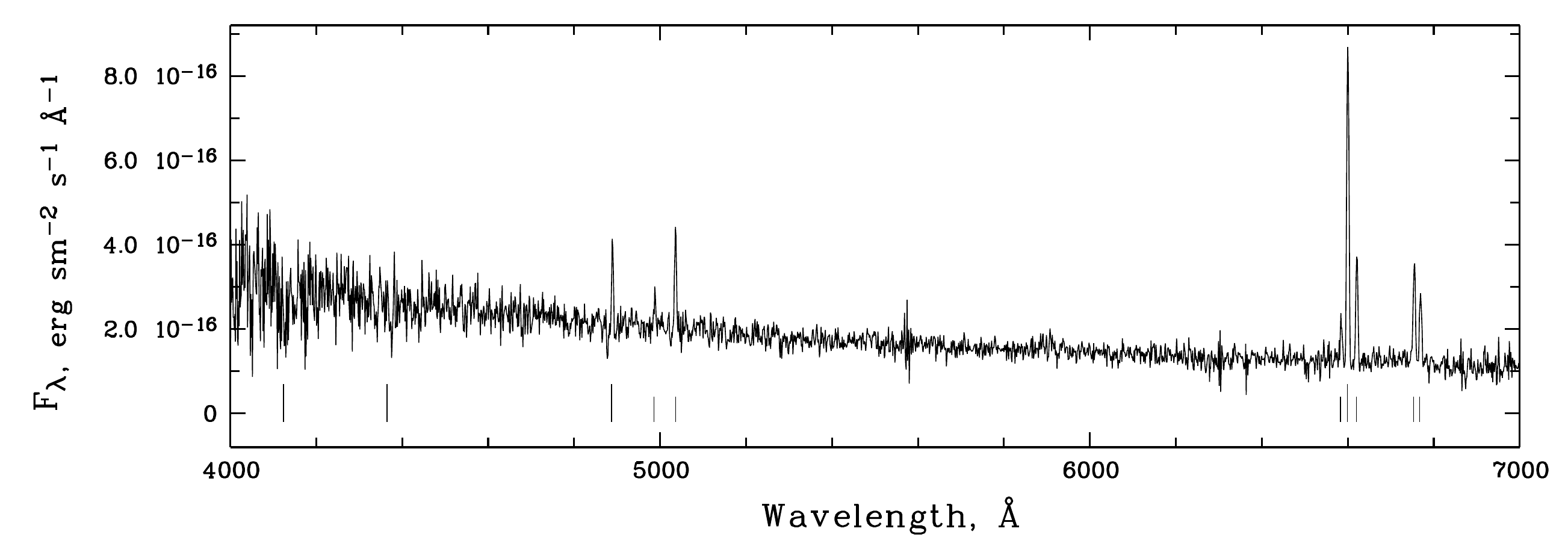}
\caption{Optical spectrum of the X-2 stellar environment and nebula. The long dashes mark the hydrogen lines H$_\delta$, H$_\gamma$ (in absorption) and H$_\beta$, H$_\alpha$ (in emission). Forbidden lines [O\,III]~4959,5007,  [N\,II]~6548,6583 and [S\,II]~6716,6731 are shown with short dashes.}
\label{fig:ulx2_spec}
\end{center}
\end{figure*}

\noindent The relatively poor seeing of 2\farcs2 during our spectral observations did not allow us to separate sources s1 and s2 in the X-2 position error box from the environment. Thus, the acquired data were used only to estimate the reddening from the Balmer decrement in the nebula assuming case B photoionization \citep{Osterbrock2006}. The obtained optical spectrum is shown in Figure~\ref{fig:ulx2_spec}. The nebula emission lines are clearly seen in the spectrum, the blue continuum and the weak absorption lines marked in the blue part of the spectrum belong to the young stellar population. Having measured the observed ratio of the emissions H$_\alpha$/H$_\beta$, we obtain the interstellar extinction value $A_V = 0.4 \pm 0.3$\,mag ($E$($B-V$)=0.13$\pm$0.10 with $R_V=3.1$). Such a large uncertainty in the reddening value is associated with a large uncertainty in the depth of H$_\beta$ absorption which is presumably from background stars. Note, that lower extinction limit only slightly exceeds the Galactic value $A_V = 0.05$\,mag \citep{Schlafly2011}.

\begin{figure*}
\centering
\includegraphics[angle=0,scale=0.65]{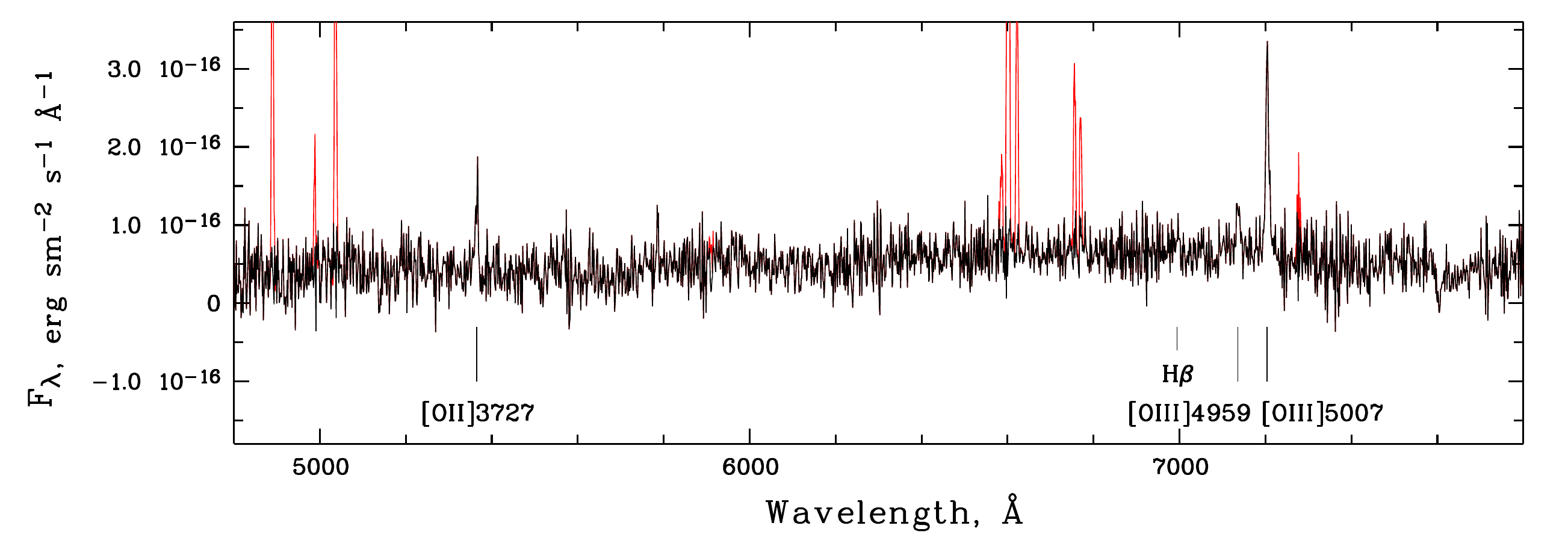}
\caption{X-3 optical spectrum obtained with SCORPIO-2 using a VPHG940@600 grating; the resolution is approximately 6\AA. The position of the H$_\beta$ and oxygen lines ([OII] 3737, [OIII] 4959,5007 corresponds to the redshift $z = 0.4391\pm0.0010$. The red color shows the spectrum before the subtraction of the lines of the background nebula located in NGC 4536.}
    \label{fig:ulx3_spec} 
\end{figure*}

\noindent The optical counterpart of X-3 is an object sufficiently isolated from bright sources, which makes it easy to extract its spectrum. However, even in this case, the spectrum of the object is contaminated by the narrow emissions from a nebula surrounding a star-forming region located approximately $8\arcsec$ from the object. After subtracting the nebula contribution, we found several emission features in the residual spectrum which were identified as oxygen lines [OII]\,3737, [OIII]\,4959, [OIII]\,5007, and weak H$_\beta$ shifted at $z = 0.4391\pm0.0010$\footnote{Statistical errors in the wavelength measurements of the three oxygen lines observed in the AGN spectrum (about 40~km\,s$^{-1}$), possible off-centre position of the target in the slit (up to 45~km\,s$^{-1}$), and the systematic shift of narrow oxygen lines known for some AGNs relative to other lines (up to $\approx200$~km\,s$^{-1}$; \citealt{Boroson2005}) are taken into account for uncertainty calculation of the redshift. The resulting redshift error is obtained as the sum of these three uncertainties.}. Thus, it is clear that X-3 is in fact an active galactic nucleus (AGN) that is visible through NGC\,4536. The optical spectrum of X-3 is shown in Figure~\ref{fig:ulx3_spec} in the 4800--7800~\AA\ range, where all detected object lines are located. We also measured H$_\alpha$/H$_\beta$ ratio in the nebula to estimate the minimal reddening on the line of sight for the AGN in the gas of the nebula and the interstellar medium in front of it. The ratio turned out to be equal to $2.80\pm0.17$; within the uncertainties it corresponds to Galactic extinction $A_V=0.05$ \citep{Schlafly2011} adopted for correction of the AGN magnitudes.

\subsection{X-ray Analysis}
The {\it XMM-Newton} data reduction was carried out with {\scshape sas} (Science Analysis Software) version 18.0.0 \citep{Gabriel2004}. Prior to analyses, re-calibration was performed by using the latest CCF (Current Calibration Files). {\scshape epchain} and {\scshape emchain} tasks were used to obtained calibrated event files of EPIC pn and MOS cameras, respectively. Both event files were filtered in the 0.2 - 12 keV energy band by using the standard event selection expressions within {\scshape evselect} task. The energy spectra of the ULX candidates were extracted by selecting circular regions with a radius of $15\arcsec$ around the sources. For the background spectra, the same size circular regions were used on a source-free region and on the same chip with the sources. X-3 was in the chip gap in the pn image. Therefore only the MOS data were used in the spectral analysis of X-3.\\

\noindent The spectral analysis of the X-ray data was performed using {\scshape xspec} version 12.11.0 \citep{Arnaus1996}. First the {\scshape grppha} task in {\scshape ftools} package \citep{Blackburn1995} was used to group the energy spectra of the sources to have a minimum of 20 counts per bin. Two absorption components (TBABS model in {\scshape xspec}) are included; one was fixed to the Galactic value \citep[$0.02\times10^{22}$ cm$^{-2}$,][]{Dickey1990} and the other was left free to estimate the intrinsic absorption. Powerlaw (PL), disk blackbody (DISKBB) and PL+DISKBB models were used to fit the spectra of the sources. The best-fit model parameters are given in Table~\ref{tab:best-fit-param}. The unabsorbed flux values were estimated using the CFLUX convolution model in {\scshape xspec} in the energy range of 0.3 - 10 keV and the luminosity values were calculated at the adopted distance. In addition, we also extracted the optical spectral index, $\alpha_{ox}$ as defined by \citet{Tananbaum1979}. This index depicts a UV-X-ray correlation involving the 2 keV and 2500 \r{A} luminosities respectively, and has been well documented for AGN \citep{Vagnetti2013,Lusso2010}.  In a recent study, \citet{Sonbas2019} demonstrated the correlation for ULXs and further showed that the slope of the correlation is distinctly different for ULXs and AGNs, a distinguishing feature they suggested as a potential identifier for these different class of sources.\\
\\
\begin{table*}
	\centering
	\caption{The best-fit model parameters for sources X-2 and X-3.}
	\label{tab:best-fit-param}
	\begin{tabular}{lccccccr} 
		\hline
Source & Model & $N_{H}$ $^{1}$ & $\Gamma$ & $T_{in}$ & $\chi^{2}$ (dof) & $L_{X}$ $^{2}$  \\
\\
  & & $10^{22}$ cm$^{-2}$ & & keV &  &  erg s$^{-1}$ \\
  \hline
& & & & & & \\
\vspace{0.1cm}
X-2 & PL & $0.17_{-0.01}^{+0.01}$ & $2.98_{-0.11}^{+0.11}$ & & 101.07 (102) & $7.18_{-0.51}^{+0.51} \times 10^{39}$\\
\vspace{0.1cm}
& DISKBB &<0.02 & & $0.48_{-0.01}^{+0.01}$ & 115.08 (102) & $2.59_{-0.16}^{+0.16} \times 10^{39}$\\
\vspace{0.1cm}
& PL+DISKBB & $0.14_{-0.01}^{+0.01}$ & $2.78_{-0.12}^{+0.12}$ &  $0.27_{-0.04}^{+0.02}$ & 100.86 (100) & $6.04_{-0.40}^{+0.41} \times 10^{39}$\\
& & & & & & \\
\vspace{0.1cm}
X-3 & PL & $0.30_{-0.22}^{+0.34}$ & $1.16_{-0.24}^{+0.32}$ &  & 16.61 (11) & $5.31_{-1.71}^{+2.05} \times 10^{43}$\\
& DISKBB & <0.50 & & $2.44_{-0.24}^{+0.20}$ & 16.90 (11) & $4.54_{-1.77}^{+2.60} \times 10^{43}$\\
\hline
	\end{tabular}
	\\ Notes: (1) Intrinsic absorption values (hydrogen column densities) toward the sources which were calculated by using TBABS model. (2) The Luminosity values were calculated in the 0.3 $-$ 10 keV energy range. The adopted distance of the host galaxy was used for X-2 while calculating the luminosities. The distance for X-3 was calculated from the redshift value obtained from the optical spectroscopy of the optical counterpart.
\end{table*}
\begin{table*}
\centering
\caption{A summary of properties of the optical counterpart candidates.}
\begin{tabular}{ccccccccccccc}
\hline
Optical counterparts &  & M$_{V}$ & (B-V)$_{0}$& OT & Age & log(F$_{X}$/F$_{V}$) & $\alpha_{ox}$ & log(l$_{2500}$)   \\
&  & (1) & (2) & (3) & (4) & (5) & (6) & (7)  \\
\hline
X-2   & s1$^a$  &   -6.2  & -0.3 & O$-$B giant/supergiant & 4$-$10 & 81 & -1.05$\pm$0.01 & 23.58$\pm$0.04 \\
   & s2 & -4.9    & - & K$-$M supergiant & $>$20 & 315 & - & - \\
X-3   & &   -19.08  & 1.54 & AGN & -  & 16 & -0.68$\pm$0.05  & 27.06$\pm$0.11 \\

\hline
\end{tabular}
\\ Notes: $^a$ assuming s1 is a point-like source (see text for details). (1)Absolute magnitudes were obtained from ACS/WFC3 data with adopted distance of 15.8 Mpc. (2) Color value obtained from F438W-F555M (B-V) filters. (3) Object types. (4) Age of the stars in Myr. (5) The F$_{X}$/F$_{V}$ ratios, (6) X-ray-UV correlation cast in terms of the optical spectral index, $\alpha_{ox}$. (7) Luminosity at 2500\AA\ in erg\,s$^{-1}$\,Hz$^{-1}$  

\label{T:tab4}
\end{table*}

\section{Results and Discussion}
\subsection{Optical}

As already noted in Section~\ref{sec:phot}, we cannot unambiguously conclude whether s1, one of the two possible X-2 optical counterparts, is point-like or an extended source. In the first case the source emission is likely due to a combination of both the accretion disk and the donor star. If we assume that the donor dominates, its spectral type is estimated as O or early B giant or supergiant (up to Iab). This is based on the intrinsic color $(B-V) = -0.39 \pm 0.21$\footnote{The intrinsic $(B-V)$ color was calculated using PySynphot from F438W and F555W magnitudes.} and the absolute magnitude $M_{V} = -6.2 \pm 0.4$\,mag, where errors take into account the statistical errors, the statistical and systematic accuracy of the  distance modulus from \cite{Hatt2018} and reddening uncertainty. Theoretical stellar parameters were taken from \cite{Fitzgerald1970,Straizys1981}. On the other hand, both $M_V$ and color value are characteristic of most bright blue ULX counterparts \citep{Tao2011} dominated by radiation from a hot supercritical disk as proposed in \cite{Vinokurov2018}. If s1 is an extended source, then it is most likely that it is a star cluster or an association. X-2 can be associated with one of the members of this cluster, and its brightness will be significantly lower than the total brightness of the cluster ($M_{V} \approx -6.2$). In this case, the optical counterpart of X-2 will be among the weak sources in the distribution of ULXs by absolute magnitudes \citep{Vinokurov2018}.\\ 

\noindent The second optical counterpart candidate, s2, is detected as a point-like source with reliably measured brightness only in the F555 and F814 filters. In blue filters, the source is below the detection threshold with a signal-to-noise ratio S/N = 3. Its intrinsic $(V-I_c)=1.7\pm0.4$; the color, calculated from F555W-F814W, and the absolute magnitude $M_V = -4.9\pm0.4$, correspond\footnote{$(V - I_c)$ colors for red supergiants were calculated using ATLAS9~\citep{Castelli2003} stellar atmosphere models for solar abundance} from late K to early M supergiants close to luminosity class Ib. Donors of this type were previously detected in some ULXs by spectral methods based on absorption features in the spectra \citep{Heida2019,Lopez2020}.\\

\begin{figure}
\begin{center}
\includegraphics[angle=0,scale=0.40]{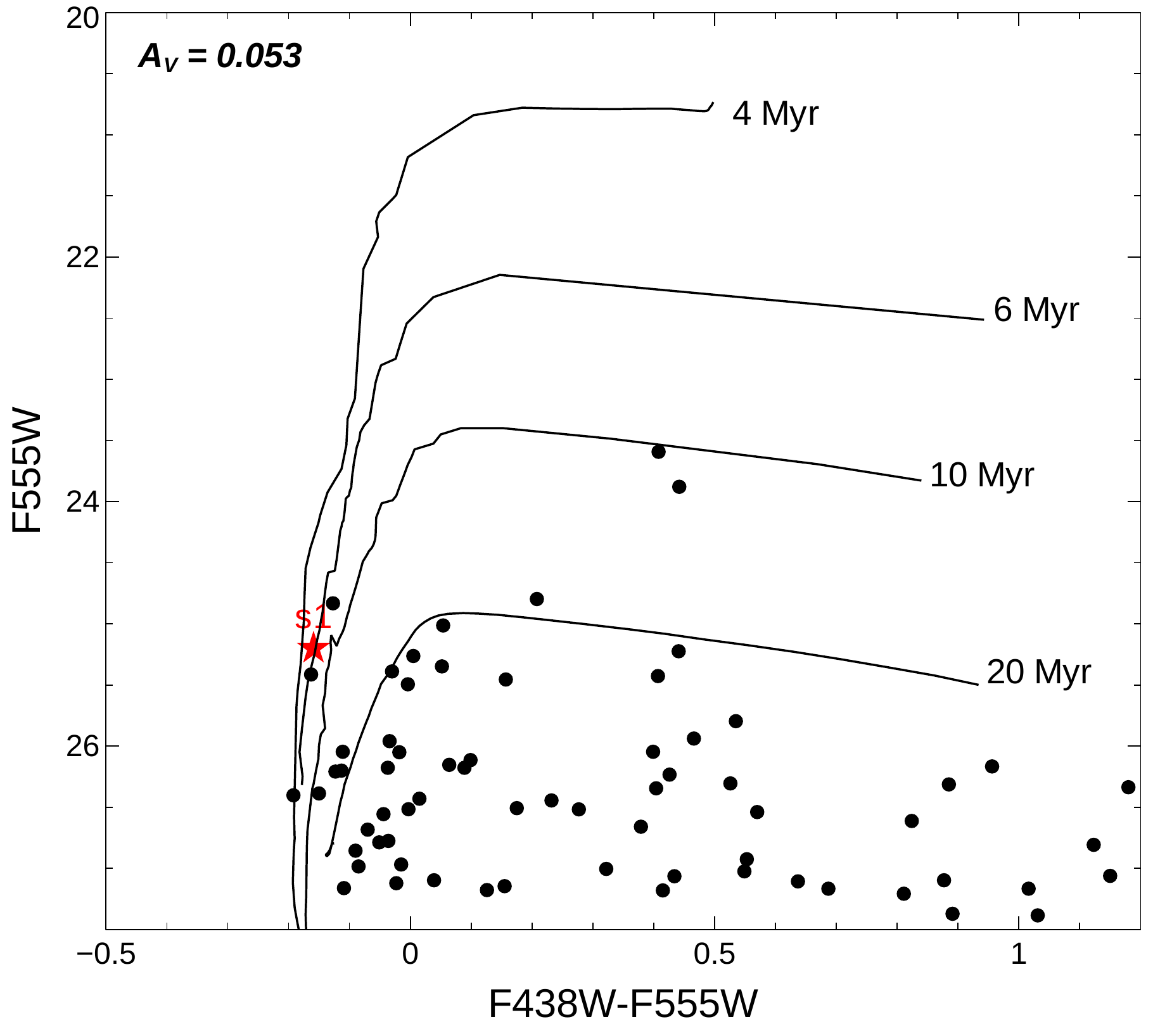}
\includegraphics[angle=0,scale=0.40]{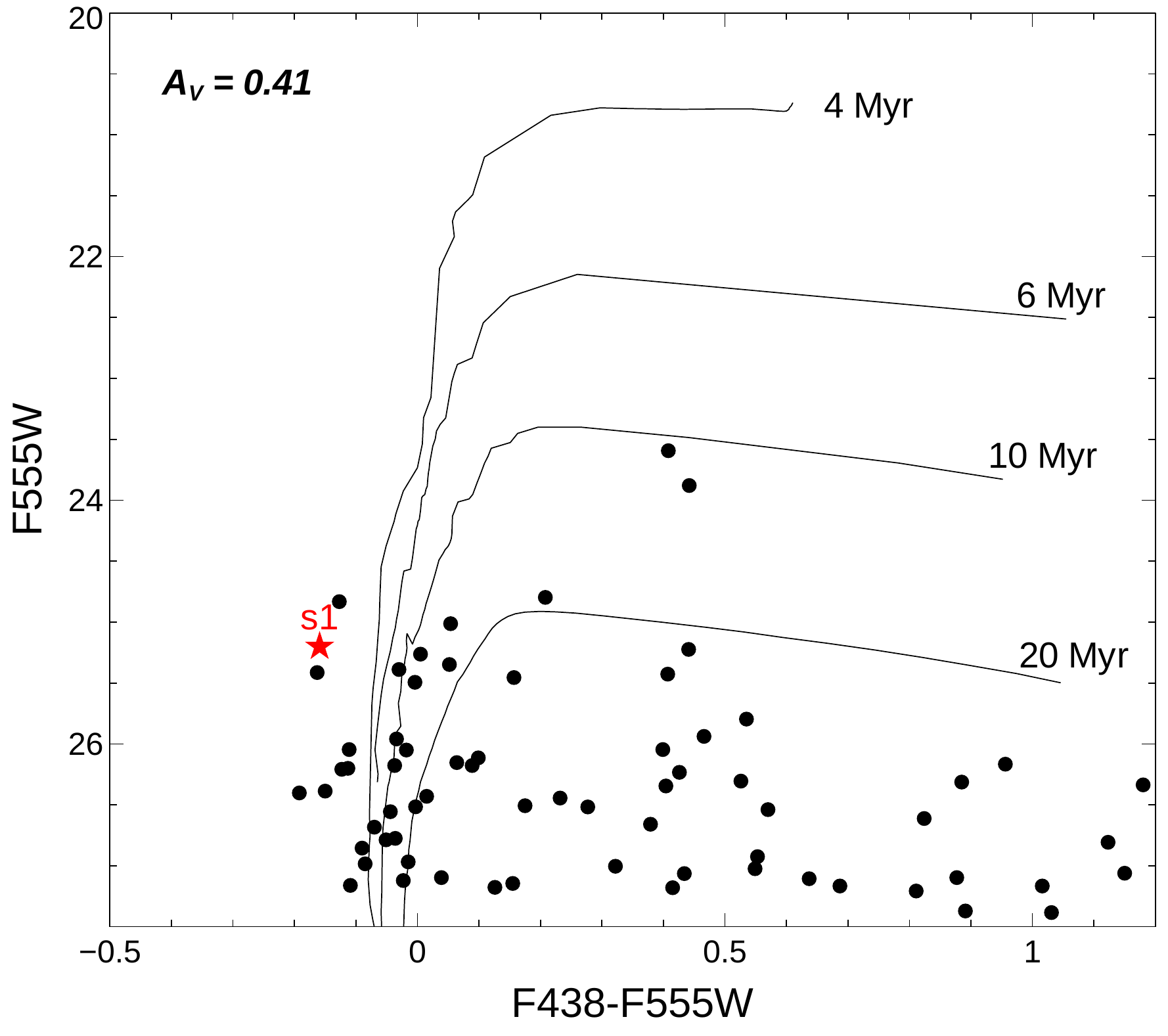}
\caption{The {\it HST}/WFC3 CMDs for s1. The source is shown with red star and field stars within 3\arcsec around X-2 are shown with black dots. The Padova isochrones with different star ages were corrected for two different extinction values ($A_{V}$= 0.053 mag and $A_{V}$= 0.41 mag).}
\label{F:cmd_s1}
\end{center}
\end{figure}

\begin{figure}
\begin{center}
\includegraphics[angle=0,scale=0.40]{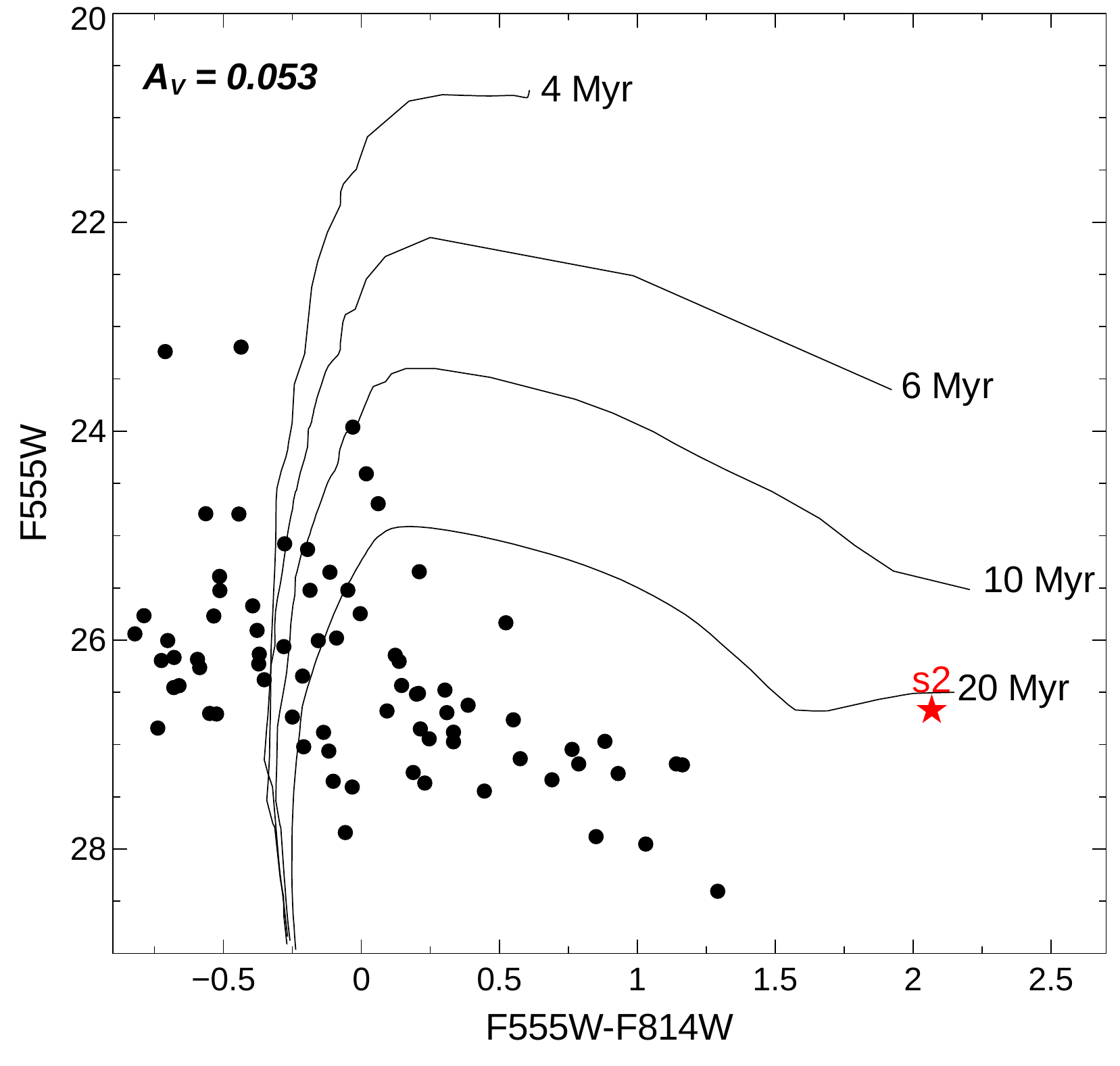}
\includegraphics[angle=0,scale=0.40]{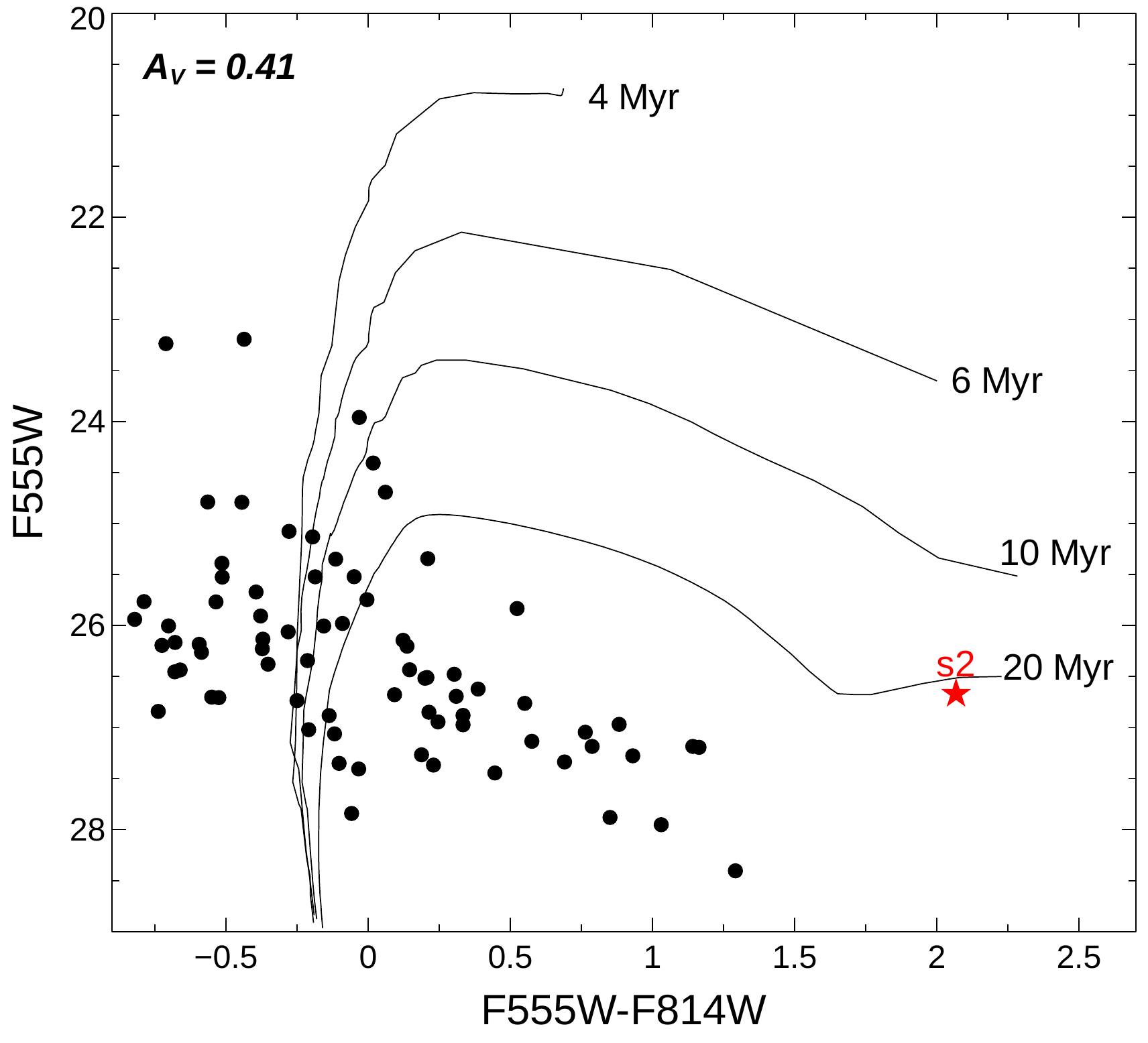}
\caption{The {\it HST}/WFC3 CMDs for s2. The source is shown with red star and field stars within 3\arcsec around X-2 are shown with black dots. The Padova isochrones with different star ages were corrected for two different extinction values ($A_{V}$= 0.053 mag and $A_{V}$= 0.41 mag).}
\label{F:cmd_s2}
\end{center}
\end{figure}

\noindent If the optical emission is assumed to be dominated by the donor star, it is possible to constrain the age of the donor star by comparing the source with theoretical isochrones. We therefore constructed color-magnitude diagrams (CMDs) for the optical candidates of X-2 in order to estimate the age of the sources. F555W (roughly corresponds to Johnson $V$ band) versus F438W-F555W $(B-V)$ diagram was used for s1. Since s2 was not detected in F438W filter, the CMDs for this source were obtained using F555W versus F555W-F814W $(V-I)$. The PARSEC (Padova and Trieste Stellar Evolution Code) isochrones\footnote{http://stev.oapd.inaf.it/cgi-bin/cmd} for HST/WFC3 photometric system were adopted (Bressan et al. 2012). The metallicity was taken as solar value while constructing these isochrones. Two CMDs were obtained for both sources by using extinction values $E(B-V) = 0.017$ (Galactic value) and $E(B-V) = 0.13$ (obtained from optical spectra) with the distance modulus 30.99 mag \citep{Hatt2018}. The CMDs for the optical counterpart candidates of X-2 (and field stars) are shown in Figure~\ref{F:cmd_s1} and \ref{F:cmd_s2}. As can be seen in the Figure~\ref{F:cmd_s1}, the CMD obtained using Galactic extinction gave better result for s1. Therefore we estimated the age values for s1 using the upper panel in Figure~\ref{F:cmd_s1}. In the case of s2, both extinction values yielded a similar result. The estimated age values are given in Table~\ref{T:tab4}.\\
\\

\noindent The spectrum of X-3 convincingly shows that this source is indeed an AGN with a redshift of $z = 0.4391\pm0.0010$, and not an X-ray source in NGC\,4536 galaxy. This serves as an illustration of the importance of optical observations, especially spectroscopy, as a valuable asset in clarifying the nature of X-ray sources. The estimated parameters for the optical counterpart of X-2 and X-3 are given in Table~\ref{T:tab4}.\\

\subsection{X-rays}
The X-ray to optical flux ratios ($F_{\mathrm{X}}/F_{\mathrm{opt}}$) were calculated \citep{Aird2010,Avdan2016} and were found to be 81, 315 for s1 and s2 (of X-2), and 16 for X-3, respectively. The ratios for s1 and s2 are similar to the range for the known ULXs. The ratio for X-3 is in agreement with the range that is derived from the majority of the AGNs (0.1 - 10) \citep{Aird2010} and also below the observed range for ULXs \citep{Avdan2016,Fabrika2021}. The extracted values of $\alpha_{ox}$ are listed in Table~\ref{T:tab4}. We note that the $\alpha_{ox}$ for X-2 agrees very well with those reported by Sonbas et al. (2019) for a sample of well-known ULXs. The $\alpha_{ox}$ for X-3, on the other hand, does not follow the ULX correlation but instead is in excellent agreement with values typically associated with AGN \citep{Vagnetti2013,Mcalphine2011,Lusso2010,Sonbas2019}. This difference in the $\alpha_{ox}$ alone suggests that X-3 is not a ULX but is instead a background AGN.\\ 

\begin{figure*}
\centering
\begin{subfigure}[]{}
\includegraphics[angle=0,scale=0.35]{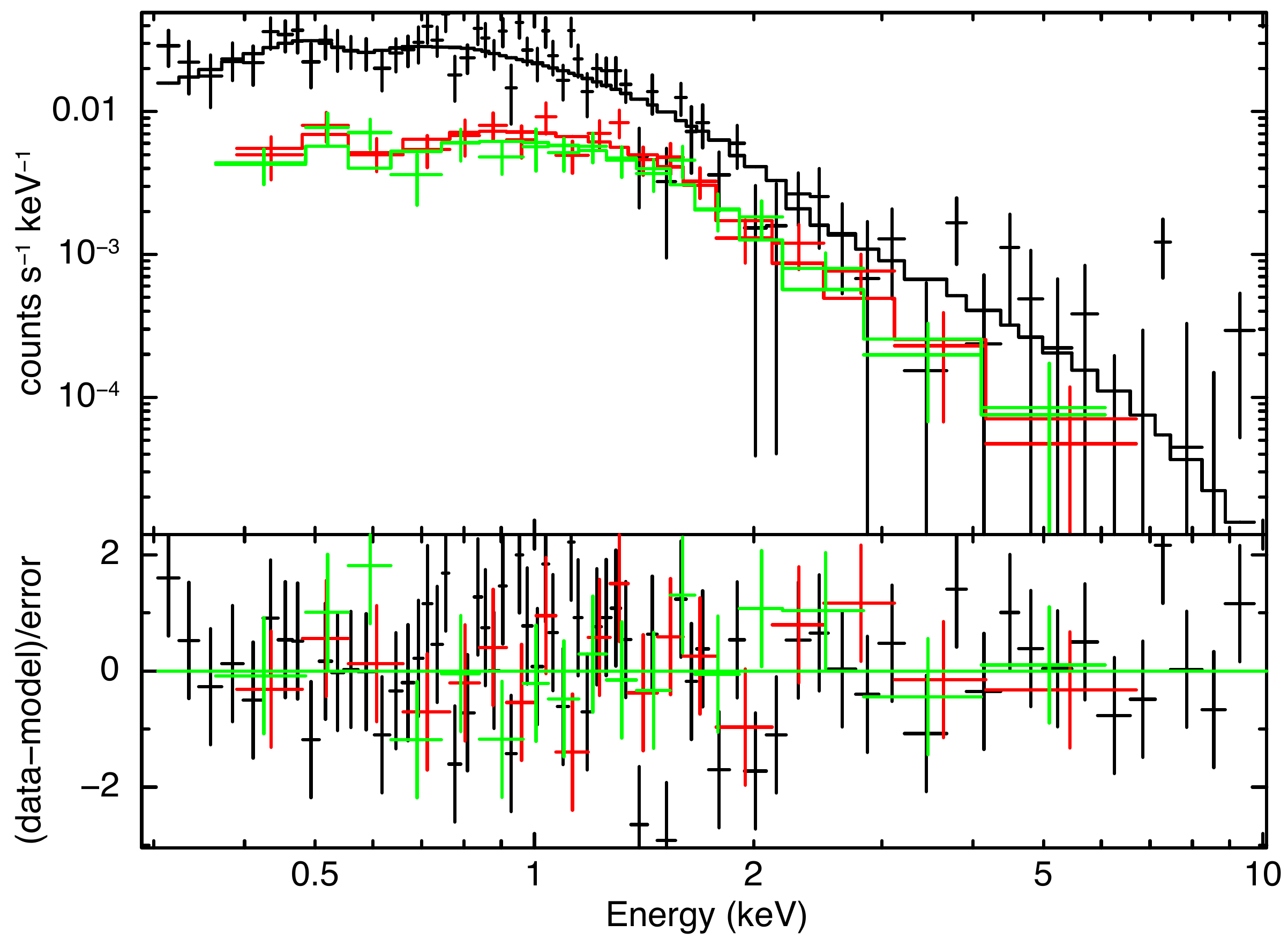}
\end{subfigure}
\begin{subfigure}[]{}
\includegraphics[angle=0,scale=0.35]{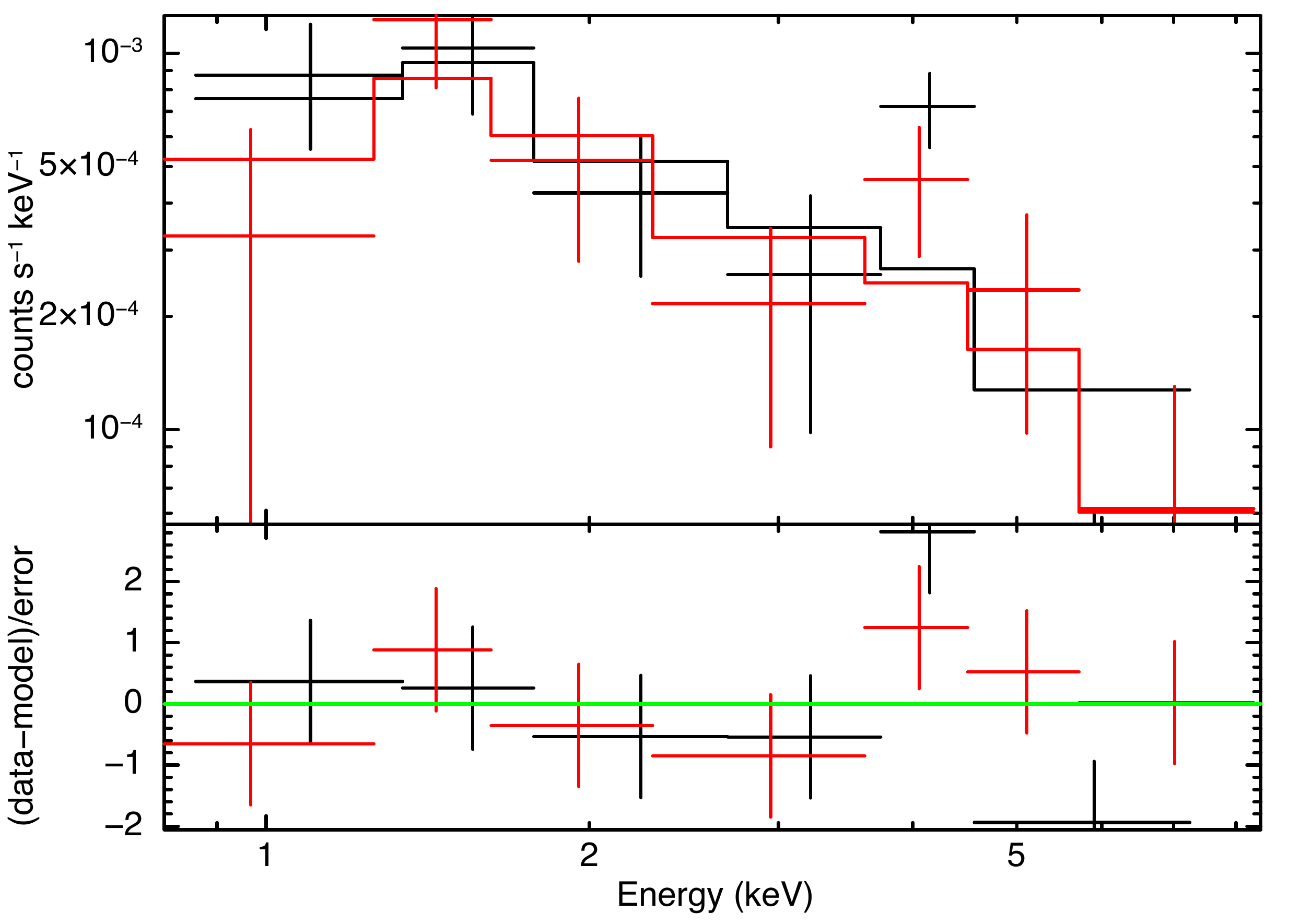}
\end{subfigure}
    \caption{(a) The {\it XMM-Newton} pn (black), MOS1 (red) and MOS2 (green) energy spectrum of X-2. The fitted model is PL+DISKBB. (b) The {\it XMM-Newton} MOS1 (black) and MOS2 (red) energy spectra of X-3. The fitted model is PL.}
    \label{fig:energy_spectra}
\end{figure*}

\noindent Our analysis of the X-ray spectra for the two sources is somewhat less firm primarily because of the low photon statistics but nonetheless we report the results here for completeness. The best-fit model parameters are given in Table~\ref{tab:best-fit-param}. As can be seen in Table~\ref{tab:best-fit-param}, for X-2 PL+DISKBB and for X-3 PL model yielded a slightly better fit although arguably the difference in the $\chi^2$ is minimal. The energy spectra of the sources with the best-fit models are displayed in Figure~\ref{fig:energy_spectra}. It is noteworthy that X-2 fit parameters suggest a significant thermal emission component while X-3 tends to favor a larger non-thermal component (with a spectral index $\Gamma$ = 1.16 $\pm$0.28), as would be expected in the case of an AGN source. This index is relatively low compared to what one might expect based on a sample of the general AGN population.  However, we note that this value is consistent with the value given by \citet{Liu2016},  in which they performed a spectral analysis of a number of low-redshift, low-luminosity X-ray- obscured type 2 AGN (see their Table 6).  We also mention the work of \citet{Lehmer2021}, who studied the X-ray population in NGC 4536. Their source $\#$16 is our X-3 source,  for which they report a spectral index of 1.49$\pm$0.93,  in agreement with our result.  They list the column density $N_{H}$ as (9.7$\pm$11) x 10$^{21}$ cm$^{-2}$: The very large uncertainty suggests that either the absorption due to the host galaxy is negligible or their model fit and/or their spectrum statistics are inadequate to constrain the column density. 
\\ 
\section{Summary and Conclusions}
We have deployed archival {\it Chandra}, {\it XMM-Newton}, and {\it HST} data to search for optical counterparts of two previously catalogued ULX candidates in NGC\,4536. We applied an astrometric correction to extract accurate coordinates of potential optical counterparts. Also, the optical spectra were used to extract the redshift from the observed oxygen lines. The X-ray and optical fluxes were used to determine the $F_{\mathrm{X}}/F_{\mathrm{opt}}$ ratios, and the luminosities at 2 keV and 2500 \r{A} respectively were used to calculate the optical spectral indices for both sources. The available X-ray spectra were fitted with appropriate models in order to ascertain whether the emissions were thermal, non-thermal or exhibited features consistent with the presence of both components. We summarize our main findings as follows:
\begin{itemize}
\item {The flux ratios are in a range that is consistent with the ratios observed for ULXs and AGN}
\item {The optical spectral index for X-2 is consistent with that found for ULXs as reported by Sonbas et al. (2019). However, the index for X-3 is inconsistent with values for ULXs and is, instead, in agreement with the index found for low-z AGN \citep{Vagnetti2013,Lusso2010,Sonbas2019}}
\item {The X-ray spectrum of X-2 is well fitted with a disk blackbody model suggestive of predominantly thermal emission. X-3, on the other hand,  presents a slightly better fit with a \textit{powerlaw}, with a spectral index of $\Gamma$ = 1.16 $\pm$0.28, indicating a significant non-thermal component in the emission.}
\item {The optical spectrum obtained with the BTA 6m telescope for X-3 shows a number of redshifted oxygen lines; the extracted redshift (z = 0.4391$\pm$0.0010) is higher then the host galaxy, a clear indication that this source is a background AGN. No such shift is detected for X-2.}
\item {Two candidates have been identified as optical counterparts for X-2, one of which is a blue source (s1) and the other is a red one (s2). The star-like nature of s1 raises some doubts i.e., we cannot rule out that it is a compact star cluster. In that case, X-2 could be one of the members of this cluster, and its brightness then would fall in the low-end of the optical range for ULXs. If, on the other hand, s1 is a point-like source and its measured extent is due to a very close and complex environment, then the reddening-corrected absolute magnitude and color place it in the category of an O-B star (with luminosity class III to Iab). The absolute magnitude and color of the second candidate (s2) correspond to a red supergiant. Based on CMDs, we estimate ages of both s1 and s2 for the case of donor star domination in their optical emission as 4$-$10 Myr and $>$20 Myr, respectively.}

\end{itemize}
\section*{Acknowledgements}
This research was supported by the Scientific and Technological Research Council of Turkey (TÜBİTAK) through project number 119F334. Observations with the SAO RAS telescopes are supported by the Ministry of Science and Higher Education of the Russian Federation. The renovation of telescope equipment is currently provided within the national project "Science and universities". Optical research was supported by the Russian Science Foundation (project no. 21-72-10167 ULXs: wind and donors). 

\section*{Data Availability}
The scientific results reported in this article are based on archival observations made by the Chandra and XMM-Newton Observatories. This work has also made use of observations made with the NASA/ESA Hubble Space Telescope, and obtained from the data archive at the Space Telescope Science Institute. These data are freely available from the appropriate archives. Additional optical data were acquired with the use of the BTA 6m telescope located at the Special Astrophysical Observatory (SAO). The ground-based optical spectra will be made available upon reasonable request.



\bibliographystyle{mnras}
\bibliography{export-bibtex.bib}{} 





\bsp	
\label{lastpage}
\end{document}